\documentclass{article}

\usepackage{microtype}
\usepackage{graphicx}
\usepackage{subfigure}
\usepackage{booktabs}
\usepackage{amsmath}
\usepackage{xurl}
\usepackage{hyperref}

\newcommand{\SYS}{HiState}
\newcommand{\Title}{\SYS{}: Hierarchical Prefix Cache Management for Hybrid LLM Serving in Agentic Workloads}

\usepackage[accepted]{mlsys2025}
\makeatletter
\renewcommand{\Notice@String}{Preprint.}
\makeatother
\hypersetup{pdfsubject={Preprint}}

\AddToHook{env/figure/begin}{\pdfrunninglinkoff}
\AddToHook{env/figure/end}{\pdfrunninglinkon}
\AddToHook{env/figure*/begin}{\pdfrunninglinkoff}
\AddToHook{env/figure*/end}{\pdfrunninglinkon}

\mlsystitlerunning{\Title{}}

\begin{document}
\twocolumn[
\mlsystitle{\Title{}}

\mlsyssetsymbol{equal}{*}
\begin{mlsysauthorlist}
\mlsysauthor{Zaifeng Pan}{equal,ucsd}
\mlsysauthor{Lucas Wang}{equal,ucsd}
\mlsysauthor{Xinwei Qiang}{ucsd}
\mlsysauthor{Yanbo Zhou}{ucsd}
\\
\mlsysauthor{Chang Chen}{ucsd}
\mlsysauthor{Zhengding Hu}{ucsd}
\mlsysauthor{Yue Guan}{ucsd}
\mlsysauthor{Yufei Ding}{ucsd}
\end{mlsysauthorlist}
\mlsysaffiliation{ucsd}{University of California, San Diego, CA, USA}
\mlsyscorrespondingauthor{Zaifeng Pan}{zapan@ucsd.edu}
\mlsyskeywords{Hybrid Model, KV Cache, Agent Serving}

\vskip 0.3in
\begin{abstract}
Hybrid large language models combine full-attention and recurrent linear layers to reduce inference costs.
Prefix caching can further reduce redundant computation across requests, but a recurrent checkpoint cannot recover an earlier state corresponding to a shorter prefix.
Existing hybrid prefix caching systems retain checkpoints at selected boundaries to support common reuse patterns in conventional conversations.
However, agent workloads modify and revisit earlier contexts, exposing reusable prefixes whose checkpoints were never saved or have already been evicted.
A straightforward approach is to retain more checkpoints to reduce recomputation, but their cumulative footprint can displace reusable prefixes from other sessions and cause additional cache misses.
To address these problems, we present \SYS{}, a prefix cache management system for hybrid models serving agent workloads.
Our key observation is that prefix reuse accesses all matching key-value (KV) entries but only one recurrent checkpoint, regardless of how many checkpoints are retained.
This asymmetry allows \SYS{} to retain historical checkpoints on SSD while prioritizing fast memory for KV entries: each hit retrieves only one checkpoint whose size is independent of context length.
Within this hierarchy, we propose adaptive interval control to balance checkpoint storage against recomputation, and unified eviction to coordinate checkpoint sparsification with reclamation of stale trajectories.
We implement \SYS{} on top of vLLM and evaluate it on two public coding-agent trace datasets and controlled microbenchmarks.
On real agent traces, \SYS{} achieves up to 5.73$\times$ and 5.44$\times$ speedups in p99 TTFT over vLLM-Marconi and SGLang-HiCache, respectively.
\end{abstract}
]

\printAffiliationsAndNotice{\mlsysEqualContribution}

\section{Introduction}
\label{sec:introduction}

Hybrid large language models (LLMs) combine full-attention layers with recurrent layers such as linear attention and state space models~\cite{lieber2024jamba,glorioso2024zamba,waleffe2024empirical,team2025kimi,team2026kimi,qwen2026qwen35}.
In full-attention layers, the key-value (KV) cache grows with sequence length, and each decoding step accesses the KV tensors of preceding tokens.
Recurrent layers instead update a fixed-size state that summarizes the preceding context, reducing the computation and memory required by these layers~\cite{katharopoulos2020transformers,gu2023mamba}.

Prefix caching~\cite{gao2024cost,yu2025stateful,qin2025mooncake,pan2025kvflow,liu2025lmcache} reuses previously computed model states across requests with shared prefixes, avoiding redundant prefill computation and reducing time-to-first-token (TTFT).
For full-attention models, systems such as SGLang~\cite{sglang} organize token-level KV entries in a radix tree, enabling reuse at intermediate prefix boundaries subject to cache block granularity.
However, a recurrent state represents an entire prefix and cannot be truncated to recover the state of a shorter one~\cite{pan2025marconi}.
Hybrid prefix reuse therefore requires both the attention KV cache and a recurrent-state checkpoint at or before the matching boundary.
The system must recompute any intervening tokens from the checkpoint, or the entire prefix if no earlier checkpoint is available.

\begin{figure*}
    \centering
    \includegraphics[width=0.95\textwidth]{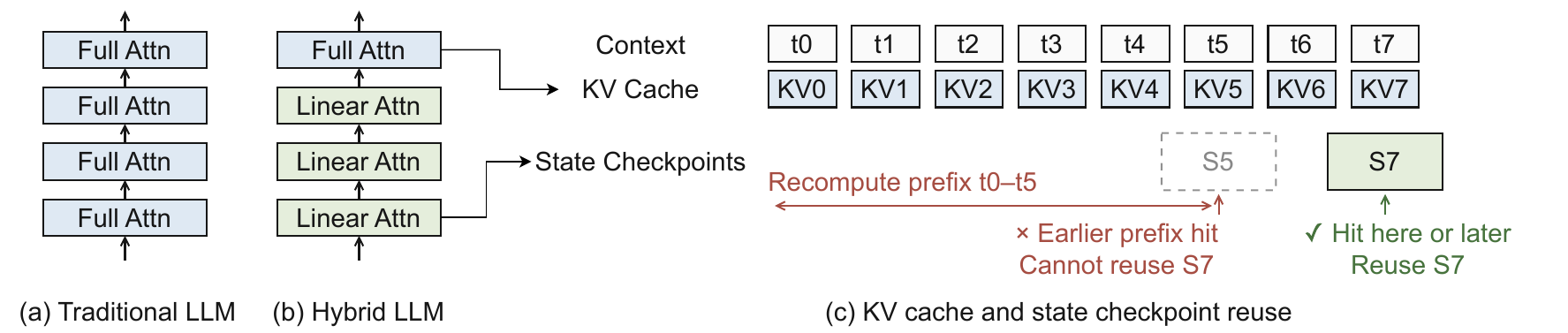}
    \caption{Model architectures and prefix cache reuse. (a) A full-attention model. (b) A hybrid model with linear-attention layers. (c) KV entries support reuse through t5, but the checkpoint S7 at t7 cannot be converted back into S5. Without S5 or an earlier checkpoint, the system must recompute t0--t5.}
    \label{fig:hybrid-cache}
\end{figure*}

Existing hybrid caching strategies exploit recurring prefix boundaries to limit checkpoint storage.
For example, Marconi~\cite{pan2025marconi} saves states at observed branching points and the last decoded token.
These choices capture common reuse patterns in question answering and conversations, where requests share a fixed system prompt or append to the preceding context.

However, agent workloads~\cite{yao2023react,yang2024swe,openclaw,langchain} introduce more diverse prefix reuse patterns through dynamic context construction.
During task execution, agent harnesses may edit earlier tool outputs, compact selected context segments, or rewind to a previous context before exploring a different execution branch~\cite{ji2025manus,langchain_context_engineering,context_for_deepagents,anthropic_context_engineering,pan2026smoothagent}.
These operations can move the prefix-reuse boundary to an internal position without a saved checkpoint, requiring recomputation even when the matching KV cache remains available.
Moreover, reuse at a previously checkpointed boundary may occur only after intervening requests have caused that checkpoint's eviction.
Supporting agent workloads therefore requires deciding both where to save checkpoints and which checkpoints to retain for later reuse.

A straightforward approach is \textit{fine-grained checkpointing}, which retains recurrent states at small, fixed token intervals~\cite{pan2025marconi}.
With an interval of $I$ tokens, fewer than $I$ tokens need to be recomputed before the matching boundary, as long as the required checkpoints and KV entries remain available.
However, a context of length $L$ requires approximately $L/I$ checkpoints, each containing the states of all recurrent layers.
For example, checkpointing Qwen3.5-9B~\cite{qwen2026qwen35} every 1,024 tokens over a 32K-token context requires approximately 1.6~GB for FP32 recurrent matrices alone, compared with 1.1~GB for its BF16 KV cache.
As concurrent sessions accumulate checkpoints, they compete with reusable KV entries for GPU and host memory.
Reducing recomputation at individual reuse positions can therefore introduce cache misses elsewhere.

To address these limitations, we propose \SYS{}, a prefix cache management system for hybrid models serving agent workloads.
Our design exploits the different access requirements of KV entries and recurrent checkpoints: resuming from a cached position requires the KV entries of all preceding tokens but only one recurrent checkpoint at that position.
Retaining more checkpoints increases the available resume positions without increasing the recurrent-state data accessed by each resumption.

Based on this observation, \SYS{} introduces \textit{asymmetric cache tiering} across GPU HBM, CPU DRAM, and SSD.
The policy prioritizes fast-memory capacity for KV entries and retains historical checkpoints on SSD.
Unlike KV restoration, checkpoint restoration transfers a fixed amount of data independent of context length and checkpoint density.
This placement uses SSD capacity to preserve historical resume positions while reducing competition with reusable KV prefixes in fast memory.

While tiering expands cache capacity, checkpoint density still determines the tradeoff between storage consumption and recomputation.
\SYS{} employs \textit{adaptive interval control} to select a checkpoint interval for each newly added context segment according to its length and runtime resource pressure.
The controller balances the expected recomputation at internal prefix hits against the storage required to retain checkpoints.

Finally, finite cache capacity requires balancing historical checkpoint retention within a session against the capacity needed by other sessions.
Preferentially retaining KV entries should not leave stale prefixes occupying space needed by active sessions.
\SYS{} introduces a \textit{unified eviction policy} that ranks KV entries and checkpoints using last-access time, reclaimable storage, and expected additional recomputation.
The policy sparsifies low-value intermediate checkpoints while allowing stale KV segments and their terminal checkpoints to be reclaimed for active sessions.

We implement \SYS{} on top of vLLM~\cite{vllm} and evaluate it on Weka~\cite{fox2026weka}, TraceLab~\cite{zhu2026tracelab}, and synthetic microbenchmarks with varying mutation frequencies.
On the real traces, \SYS{} achieves up to 5.73$\times$ and 5.44$\times$ speedups in p99 TTFT over vLLM-Marconi~\cite{vllm,pan2025marconi} and SGLang-HiCache~\cite{sglang,xie2026strata}, respectively.

In summary, this paper makes the following contributions:
\begin{itemize}
    \item We identify the checkpoint placement and retention challenges introduced by context mutation and delayed reuse in agent workloads.
    \item We design and implement \SYS{}, combining asymmetric cache tiering, adaptive interval control, and unified eviction to jointly manage KV caches and recurrent checkpoints across sessions and the memory hierarchy.
    \item We evaluate \SYS{} with real agent traces and controlled prefix reuse patterns, showing how checkpoint reuse affects request latency and workload tails.
\end{itemize}

\section{Background and Motivation}
\label{sec:background}

\subsection{Prefix Reuse in Hybrid Models}

Hybrid LLMs combine full-attention layers with recurrent layers, such as linear attention and state space models~\cite{lieber2024jamba,glorioso2024zamba,team2025kimi,qwen2026qwen35}, as illustrated in Figure~\ref{fig:hybrid-cache}(a)--(b).
A full-attention layer computes each new token's output by attending to the key-value vectors of preceding tokens.
These vectors are retained in a KV cache whose size grows with context length.
In contrast, a recurrent layer incorporates each new token into a fixed-size state that summarizes the preceding context~\cite{katharopoulos2020transformers,gu2023mamba}.
Inference updates this state in place, so preserving an earlier state for future reuse requires saving a checkpoint.

Prefix caching extends reuse across requests by retaining the cached representations of previously processed prefixes.
For full-attention models, systems such as SGLang~\cite{sglang} organize KV entries in a radix tree and match each incoming request against cached token sequences.
The system reuses the KV cache of the matching prefix and performs prefill only for the remaining tokens, subject to cache block granularity~\cite{vllm}.
Since KV entries are stored for individual tokens, a request can reuse an internal portion of a cached prefix without recomputing those tokens.

Recurrent checkpoints impose an additional constraint on prefix reuse in hybrid models.
Each checkpoint summarizes an entire prefix and cannot be truncated to recover the state at an earlier position~\cite{pan2025marconi}.
In Figure~\ref{fig:hybrid-cache}(c), a new request matches the cached tokens through t5, but the system retains only the checkpoint S7 at t7.
Although KV0--KV5 remain reusable, S7 cannot be converted back into S5, forcing the system to recompute t0--t5 to reconstruct the missing recurrent state.
More generally, the system must resume from the latest checkpoint at or before the matching position and recompute the intervening tokens.
If no such checkpoint is available, it must recompute the entire matching prefix, even when the corresponding KV cache remains available.

\subsection{Agent Context Mutation Creates Internal Hits}

Conventional chat workloads exhibit relatively stable prefix reuse patterns.
Within a session, each turn typically appends new messages to the existing conversation, allowing the next request to reuse the preceding turn's context.
Across sessions, requests often share a fixed system prompt.
Checkpoints at the system-prompt boundary and the end of the preceding turn can therefore support these reuse patterns without retaining states throughout the context.
Existing systems exploit such recurring boundaries to limit checkpoint storage.
For example, Marconi~\cite{pan2025marconi} retains states at observed branching points and the last decoded token.

\begin{figure}[t]
    \centering
    \includegraphics[width=0.9\columnwidth]{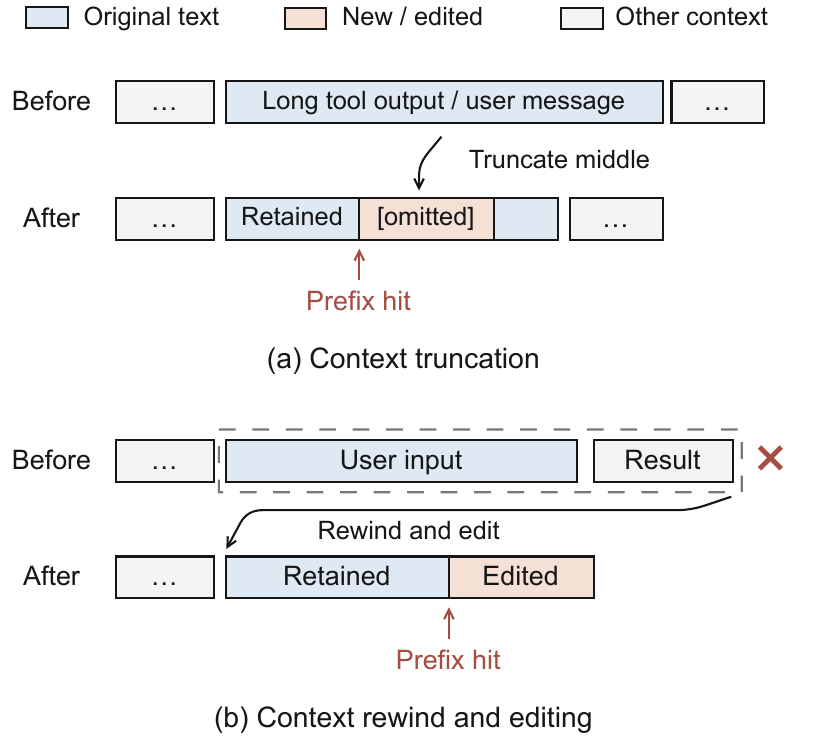}
    \caption{Context changes that create internal prefix hits. (a) Truncating a message's middle preserves a prefix ending near the omission marker. (b) Rewinding and revising an earlier input preserves a prefix ending where the edit begins.}
    \label{fig:internal-prefix-hits}
\end{figure}

Agent workloads introduce more complex context construction as task execution proceeds.
Agent harnesses may shorten existing messages to control context growth, while users or harnesses may revisit earlier steps and revise their inputs~\cite{anthropic_context_engineering,langchain_context_engineering,context_for_deepagents,pan2026smoothagent}.
These operations can move the prefix-reuse boundary inside a previously processed message, where a checkpoint may not have been retained.
Figure~\ref{fig:internal-prefix-hits} illustrates two such patterns.

\paragraph{Context truncation.}
Agent harnesses can shorten previously processed messages when constructing later requests.
For example, Codex~\cite{codex} can truncate a retained user message that exceeds the remaining compaction budget, preserving its head and tail around an omission marker.
Figure~\ref{fig:internal-prefix-hits}(a) illustrates this operation on a long message.
When the preceding context remains unchanged, the new request shares a prefix with the original request up to the first changed token near the marker.
The retained tail does not extend this prefix hit because its preceding context has changed.
If no checkpoint was saved at this internal position, the system must recompute the unchanged text between the latest earlier checkpoint and the truncation point.

\paragraph{Context rewind and editing.}
Users or agent harnesses may rewind execution to an earlier turn and revise its input before continuing.
In Figure~\ref{fig:internal-prefix-hits}(b), the original input has already produced a result when part of that input is edited and resubmitted.
For example, a coding request may retain its source code while changing a requirement at the end.
The revised request shares the preceding context and the unchanged portion of the input, so its prefix-reuse boundary falls inside the original message.
A checkpoint at the end of the original turn includes the discarded suffix and cannot initialize this revised request.
Without a checkpoint at the edit position, the system must restore an earlier checkpoint and recompute the intervening tokens.

Both operations expose reusable prefixes at internal positions that request-end and previously observed branch checkpoints may not cover.

\subsection{Inefficient Checkpoint Retention}
\label{sec:checkpoint-retention}

\begin{figure}[t]
    \centering
    \includegraphics[width=0.92\columnwidth]{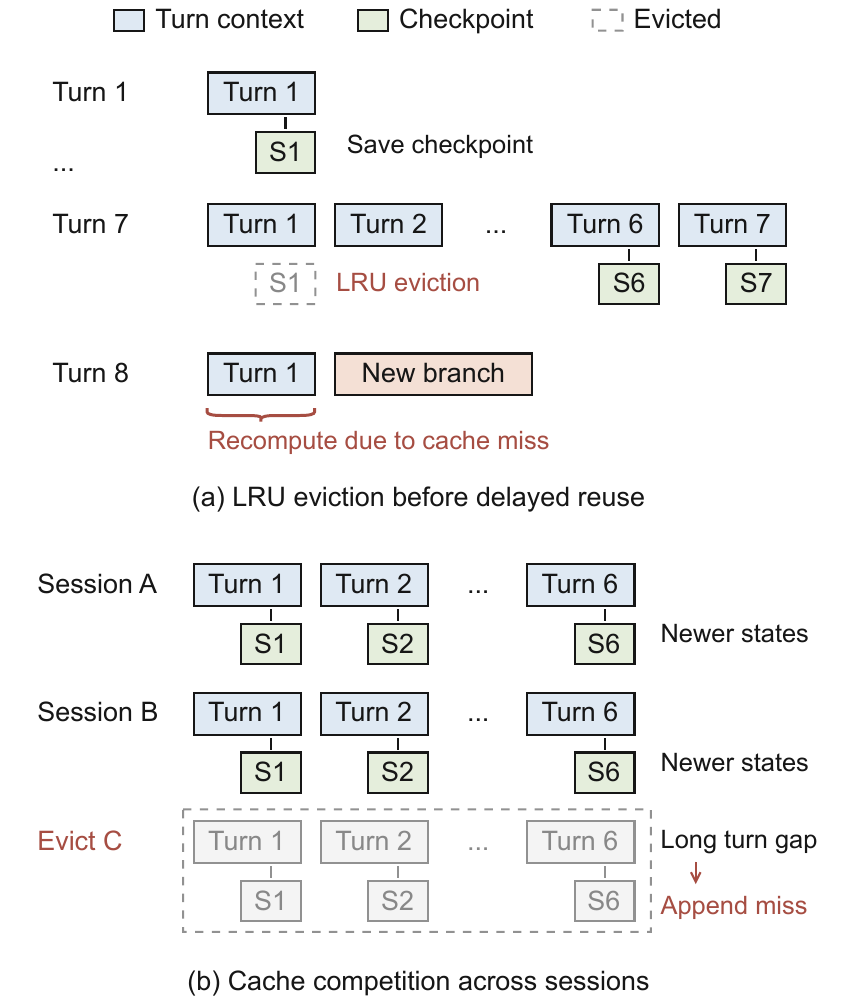}
    \caption{Checkpoint eviction before reuse under LRU. (a) S1 is saved after turn 1 but evicted under memory pressure as the session advances. Forking at that boundary in turn 8 requires recomputation. (b) While C waits for tool results after turn 6, A and B append new turns. Their newer checkpoints can displace C's latest checkpoint S6 under memory pressure, causing a miss when C appends turn 7. Each row shows a session's context, with Si denoting the checkpoint after turn i.}
    \label{fig:checkpoint-retention}
\end{figure}

Existing systems can also incur cache misses when prefix reuse occurs at turn boundaries where checkpoints were previously saved.
For example, vLLM's \texttt{align} mode~\cite{vllm} saves recurrent states at scheduler-step endpoints that align with cache blocks.
Nevertheless, we observe frequent cache misses when agent requests fork from the end of an earlier turn, even when a checkpoint was originally stored at that position.
Such retention failures can affect both historical checkpoints needed for delayed reuse and the latest checkpoint needed to continue a session.

Within a session, an agent may revisit an earlier turn only after many subsequent turns have completed.
As the session advances, each request can resume from a recent checkpoint without accessing those at earlier positions.
Under least-recently-used (LRU) eviction~\cite{vllm_prefix_caching}, these earlier checkpoints become increasingly likely to be evicted as new prefix cache entries accumulate, even though the session remains active.
When a later request forks from one of these positions, the required checkpoint may already be gone, forcing the system to recompute a prefix that was previously cached, as illustrated in Figure~\ref{fig:checkpoint-retention}(a).

Across sessions, differences in the time between turns can also make a session's latest checkpoint vulnerable to eviction.
In Figure~\ref{fig:checkpoint-retention}(b), session C finishes turn 6 and saves S6, then waits for tool results before issuing its next request.
During this interval, sessions A and B append several turns, producing checkpoints with more recent timestamps than S6.
Under memory pressure, LRU can evict S6 even though C still needs it to continue.
When C issues turn 7, its append-only request incurs a cache miss and requires recomputation to reconstruct the missing state.
Thus, a checkpoint can be the latest within its session yet appear old in the shared cache.

A straightforward way to protect historical checkpoints is to refresh the last-access time of both the matched checkpoint and all preceding checkpoints on the same prefix whenever a cache hit occurs.
This propagates the session's recent activity to its historical checkpoints, reducing their risk of eviction before delayed reuse.
However, refreshing all historical checkpoints exacerbates cache competition across sessions.
Frequent requests keep a growing set of checkpoints recent, regardless of whether those positions will be revisited.
Under memory pressure, retaining these checkpoints can displace the latest checkpoints of sessions with longer gaps between turns, causing the append misses illustrated in Figure~\ref{fig:checkpoint-retention}(b).
As requests alternate among sessions, repeated eviction and recomputation can lead to cache thrashing.
Checkpoint management must therefore balance the retention of historical states for delayed reuse with the capacity needed for other sessions to continue.

\section{System Design}
\label{sec:design}

\begin{figure}[t]
    \centering
    \includegraphics[width=\columnwidth]{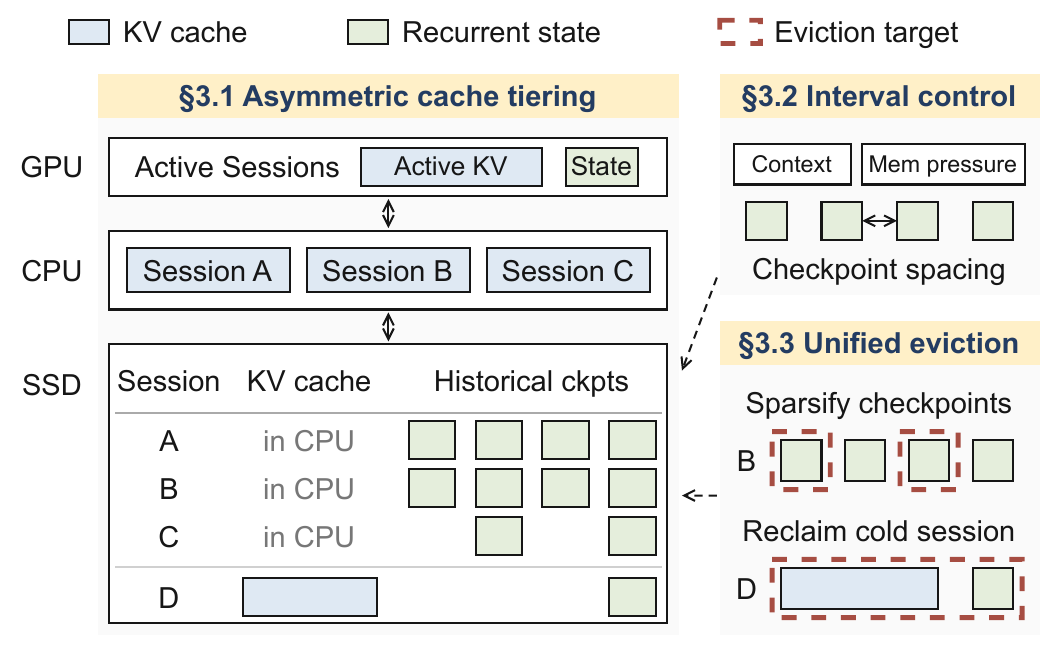}
    \caption{System overview of \SYS{}.}
    \label{fig:system-overview}
\end{figure}

Figure~\ref{fig:system-overview} shows the architecture of \SYS{}, which jointly manages KV entries and recurrent checkpoints across GPU HBM, CPU DRAM, and SSD.
The system coordinates three decisions: where to retain cached data, how frequently to create checkpoints, and which entries to evict when capacity is exhausted.

\emph{Asymmetric cache tiering} prioritizes fast memory for KV entries and stores historical checkpoints on SSD, allowing the system to retain multiple resume positions without occupying the memory needed by reusable KV prefixes.
Active KV entries and recurrent states reside on the GPU during inference.
For each incoming request, \emph{adaptive interval control} selects checkpoint spacing according to context length and resource pressure, balancing recomputation at internal prefix hits against checkpoint storage and write costs.
When space must be reclaimed, \emph{unified eviction} evaluates KV entries and checkpoints using their last access time, footprint, and recomputation cost.
It can remove intermediate checkpoints while preserving a session's KV prefix, or reclaim an older session's unshared KV entries and remaining states.
These mechanisms coordinate retention within individual trajectories and capacity allocation across concurrent sessions.

\subsection{Asymmetric Cache Tiering}
\label{sec:asymmetric-tiering}

Supporting internal prefix hits and delayed reuse requires retaining checkpoints at multiple positions along a trajectory.
However, keeping these checkpoints in GPU or host memory consumes capacity that could otherwise retain reusable KV entries.
\SYS{} addresses this competition by placing the two cache types according to their different storage and retrieval requirements.

\paragraph{Storage and access asymmetry.}
KV entries and recurrent checkpoints contribute differently to prefix reuse.
Resuming at a cached position requires the attention KV entries of all preceding tokens, but only one recurrent checkpoint at that position.
For a context of $L$ tokens and a checkpoint interval of $I$, retaining the prefix requires $Lb$ bytes of KV and approximately $(L/I)S$ bytes of checkpoints, where $b$ is the KV size per token and $S$ is the size of one checkpoint across all recurrent layers.
An individual reuse, however, requires only $S$ bytes of recurrent states, independent of both $L$ and $I$.
Retaining more checkpoints therefore increases the number of available resume positions without increasing the recurrent-state data required by each resumption.

\begin{figure}[t]
    \centering
    \includegraphics[width=\columnwidth]{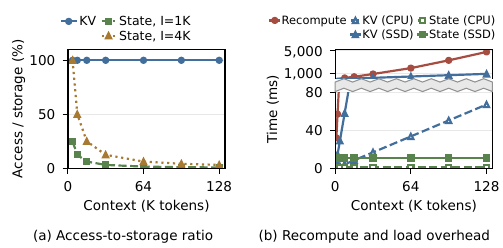}
    \caption{Cache access and restoration costs. (a) Bytes accessed per prefix reuse as a percentage of bytes retained for each cache type, evaluated at the end of the cached context. $I$ is the checkpoint interval; 1K denotes 1,024 tokens. (b) Comparison of full-prefix recomputation and KV/checkpoint loading overhead from CPU memory or SSD.}
    \label{fig:asymmetric-tiering}
\end{figure}

We quantify this distinction using the \emph{access-to-storage ratio}: the cache bytes required by a resumption divided by the bytes retained for that context.
For reuse of the entire retained context at a saved boundary, with $L$ divisible by $I$, the ratios are
\begin{equation}
    \rho_{\mathrm{KV}}=\frac{Lb}{Lb}=1,
    \qquad
    \rho_{\mathrm{state}}=\frac{S}{(L/I)S}=\frac{I}{L}.
    \label{eq:access-storage-ratio}
\end{equation}
Figure~\ref{fig:asymmetric-tiering}(a) shows that KV reuse requires all retained KV bytes, whereas the fraction of checkpoint storage accessed decreases with context length.
At 32K tokens, checkpoint intervals of 1,024 and 4,096 tokens yield ratios of 3.1\% and 12.5\%, respectively, compared with 100\% for KV.
For such a hit, keeping the KV prefix in fast memory avoids transferring all of its bytes from a slower tier, whereas keeping all checkpoints there avoids transferring only the selected checkpoint.
Retaining every checkpoint in fast memory therefore consumes capacity for states unused by that hit and can displace KV entries that must be accessed.
This asymmetry motivates prioritizing KV entries in fast memory while retaining historical checkpoints in a larger, slower tier.

\paragraph{Asymmetric restoration costs.}
The cost of retrieving data from a slower tier further supports this placement.
For $X$ bytes of cached data, loading from CPU memory requires a host-to-GPU transfer, whereas loading from SSD additionally requires an SSD-to-host transfer:
\begin{equation}
    T_{\mathrm{CPU}}(X)=\frac{X}{B_{\mathrm{H2D}}},
    \quad
    T_{\mathrm{SSD}}(X)=\frac{X}{B_{\mathrm{SSD}}}
        +\frac{X}{B_{\mathrm{H2D}}},
    \label{eq:tiering-transfer-cost}
\end{equation}
where $X=Lb$ for the KV prefix and $X=S$ for a checkpoint.
$B_{\mathrm{H2D}}$ and $B_{\mathrm{SSD}}$ denote the CPU-to-GPU and SSD-to-CPU bandwidths, respectively.
Moving the required cache data from CPU memory to SSD therefore adds $Lb/B_{\mathrm{SSD}}$ to KV restoration but only $S/B_{\mathrm{SSD}}$ to checkpoint restoration.
The former grows with context length, while the latter is independent of both context length and checkpoint density.

Figure~\ref{fig:asymmetric-tiering}(b) compares these loading costs with full-prefix recomputation for Qwen3.5-9B~\cite{qwen2026qwen35} on a single H100 (TP=1).
We estimate loading costs using BF16 KV entries, FP32 recurrent matrices, 64~GB/s CPU-to-GPU bandwidth over PCIe 5.0 x16, and 5~GB/s SSD-to-CPU bandwidth.
Loading one checkpoint takes approximately 0.8~ms from CPU memory and 10.9~ms from SSD, an additional cost of 10.1~ms.
In contrast, moving a 32K-token KV prefix from CPU memory to SSD increases its loading cost from 16.8~ms to 231.5~ms; at 128K tokens, the cost rises from 67.1~ms to 926.1~ms.
Thus, displacing KV entries to SSD incurs a growing restoration penalty, while placing historical checkpoints there adds a small, fixed cost per reuse.
For long prefixes, the measured recomputation cost substantially exceeds these loading estimates, motivating hierarchical caching to preserve reusable prefixes beyond fast-memory capacity.

These differences in access requirements and restoration costs guide the placement policy in Figure~\ref{fig:system-overview}: \SYS{} keeps active KV entries and recurrent states on the GPU, prioritizes CPU DRAM for reusable KV prefixes, and retains historical checkpoints on SSD.
Colder KV prefixes can also be offloaded to SSD when fast-memory capacity is exhausted.
On a prefix hit, the system restores the required KV entries and the checkpoint at the selected resume position.
We next describe how adaptive interval control and unified eviction regulate checkpoint density and reclaim cache capacity.

\subsection{Adaptive Interval Control}
\label{sec:adaptive-interval}

Checkpoint spacing must adapt to context length and resource pressure, even with asymmetric tiering.
A short interval consumes more cache capacity, while a long interval increases the recomputation required by an internal prefix hit.
\SYS{} selects an interval for each newly added context segment by balancing these two costs.
The final state of the segment is always checkpointed to support append reuse, regardless of the selected interval.

\paragraph{Recomputation cost.}
For a segment of $L$ new tokens and a checkpoint interval $I$, the expected recomputation cost is
\begin{equation}
    C_{\mathrm{recomp}}(I)=\sum_{t=1}^{L-1}p_t
    \left(t-I\lfloor t/I\rfloor\right),
    \label{eq:interval-recompute-cost}
\end{equation}
where $p_t$ is the probability of a prefix hit ending at position $t$, and $I\lfloor t/I\rfloor$ is the nearest checkpoint position at or before $t$.
Their distance is the number of tokens to recompute.
Without prior knowledge of future reuse positions, we assign $p_t=1/L$ to each nonterminal position.

\paragraph{Storage cost under resource pressure.}
Let $N_s$ denote the storage occupied by one state checkpoint in token equivalents.
An interval of $I$ therefore requires approximately $(L/I)N_s$ token equivalents for the segment.
To account for resource pressure, we track $T_{\mathrm{KV}}$, the estimated residence time of KV entries, and $T_{\mathrm{state}}$, the estimated residence time of checkpoints on SSD.
Both cache types must remain available for reuse, motivating similar retention horizons.
The controller therefore uses their ratio to weight checkpoint storage:
\begin{equation}
    C_{\mathrm{mem}}(I)=
    \frac{T_{\mathrm{KV}}}{T_{\mathrm{state}}}
    \frac{L}{I}N_s.
    \label{eq:interval-storage-cost}
\end{equation}
When $T_{\mathrm{KV}}/T_{\mathrm{state}}>1$, checkpoints disappear sooner than their KV entries; the higher penalty favors a larger interval to relieve SSD pressure.
Conversely, a smaller ratio lowers the penalty and permits denser checkpointing.
Both residence times are estimated online using exponential moving averages.

Then, for each new segment, the controller selects
\begin{equation}
    I^{*}=\operatorname*{arg\,min}_{I\in\mathcal{I}}
    \left[C_{\mathrm{mem}}(I)+C_{\mathrm{recomp}}(I)\right],
    \label{eq:interval-selection}
\end{equation}
where $\mathcal{I}$ is the set of supported checkpoint intervals for the segment.
Under the uniform-access approximation, the continuous solution is
\begin{equation}
    I_{\mathrm{cont}}^{*}=\sqrt{2LN_s\frac{T_{\mathrm{KV}}}{T_{\mathrm{state}}}},
    \label{eq:interval-continuous-optimum}
\end{equation}
which indicates that longer contexts or greater SSD pressure favor larger intervals and thus sparser checkpointing.

\subsection{Unified Eviction}
\label{sec:unified-eviction}

Eviction must balance checkpoint retention within a trajectory with cache allocation across trajectories.
The analysis in Section~\ref{sec:asymmetric-tiering} shows that KV entries support more reuse positions than individual checkpoints, motivating preferential retention of KV entries while sparsifying checkpoints.
However, always protecting KV entries can leave stale sessions occupying capacity needed by active sessions.
\SYS{} therefore ranks both cache types using recency, recomputation cost, and reclaimable storage.

\begin{figure}[t]
    \centering
    \includegraphics[width=0.95\columnwidth]{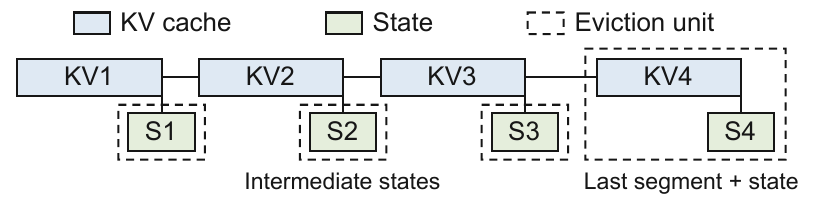}
    \caption{Eviction units along a trajectory. Intermediate states are individual units; the last KV segment and terminal state form one bundle.}
    \label{fig:eviction-units}
\end{figure}

\paragraph{Eviction units.}
An intermediate checkpoint can be evicted individually, leaving its KV entries available for reuse from other retained checkpoints.
For each trajectory, we instead group the last KV segment and terminal checkpoint into one eviction unit, since both are needed to resume at that boundary.
The candidates at each eviction step are all intermediate checkpoints and these terminal bundles across trajectories, as shown in Figure~\ref{fig:eviction-units}.

\paragraph{Staleness- and cost-aware ranking.}
Inspired by GreedyDual-Size~\cite{cao1997greedydual}, we combine recomputation cost per unit of storage with an explicit last-access timestamp:
\begin{equation}
    \begin{aligned}
        H_x &= t_x+\tau\frac{C_x}{M_x}.
    \end{aligned}
    \label{eq:unified-eviction-priority}
\end{equation}
Here, $t_x$ is the last-access time, $C_x$ is the expected additional recomputation in tokens caused by removing $x$, and $M_x$ is the storage actually reclaimed from the constrained tier in token equivalents.
The time scale $\tau$ controls the weight of recomputation cost relative to recency; the candidate with the lowest score is evicted.

We estimate $C_x$ by weighting the additional recomputation at each reuse position by its access probability.
We assume equal access probabilities across trajectories, estimate terminal-state access probabilities using exponential moving averages, and distribute the remaining probability uniformly across internal positions.
This weighting favors retaining frequently accessed terminal states while sparsifying intermediate checkpoints with low expected recomputation costs.
Affected costs are updated after eviction without changing last-access timestamps.

On a checkpoint hit, \SYS{} refreshes the timestamps of that checkpoint and all preceding checkpoints on the matched prefix.
This \emph{refresh-all} rule propagates recent activity to historical states that may be reused later, while the cost term still allows low-value intermediate checkpoints to be removed.
A stale bundle receives no such refresh and can eventually rank below checkpoints from active sessions.
Figure~\ref{fig:system-overview} illustrates these two outcomes: sparsifying session B's checkpoints and reclaiming the cold session D's KV cache and terminal state.

\begin{figure}[t]
    \centering
    \includegraphics[width=0.85\columnwidth]{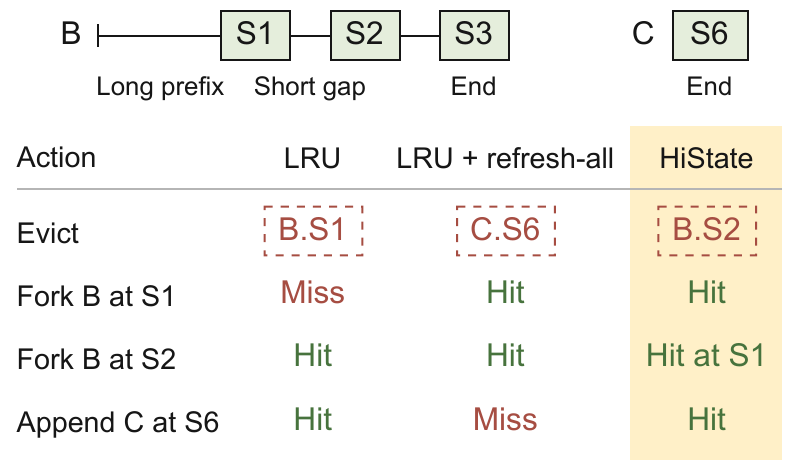}
    \caption{Checkpoint sparsification in \SYS{} compared with LRU and LRU + refresh-all. Each policy frees one checkpoint slot; reuse queries are evaluated independently. ``Hit at S1'' requires recomputation from S1 to S2.}
    \label{fig:checkpoint-sparsification-case}
\end{figure}

Figure~\ref{fig:checkpoint-sparsification-case} compares checkpoint sparsification in \SYS{} with LRU and LRU + refresh-all under the same checkpoint capacity.
Session B retains historical states S1 and S2 and its terminal state S3, while session C retains its terminal state S6.
Each policy must evict one of these four checkpoints.
LRU removes B's historical state S1 because it has not been accessed recently, causing a miss when B later forks at S1.
LRU + refresh-all keeps B's historical states recent but evicts C's terminal state S6, causing a miss when C appends.
These decisions reflect the two retention failures discussed in Section~\ref{sec:checkpoint-retention}: losing historical reuse within a session and displacing another session's latest state.

\SYS{} evicts B's intermediate state S2 in this example because it can be reconstructed from the nearby S1 at low cost.
This preserves S1, which avoids recomputing B's long prefix, and S6, which allows C to continue from its latest state.
If B instead forks at S2, it resumes from S1 and recomputes only the short gap between them.

\section{Evaluation}
\label{sec:evaluation}

\newcommand{\evaltodo}[1]{\textit{[#1]}}
\newcommand{\evalplaceholder}[2]{%
    \fbox{\parbox[c][0.78in][c]{\dimexpr#1-2\fboxsep-2\fboxrule\relax}{%
        \centering\small\textit{Result placeholder}\\[4pt]#2}}}

\subsection{Experimental Setup}
\label{sec:eval-setup}

\paragraph{Implementation and hardware.}
We implement \SYS{} on top of vLLM~\cite{vllm} and evaluate Qwen3.5-9B~\cite{qwen2026qwen35} on a single NVIDIA H100 GPU with 80~GB of HBM3.
Each system runs alone on the server with 20 Intel Xeon 8468 vCPUs.
\SYS{} uses a shared GPU cache pool for KV entries and recurrent states, comprising 2,560 blocks of 576 tokens each and occupying approximately 45~GiB.
We allocate 96~GiB of pinned CPU memory and provide 150~GiB of disk space for caching.

\paragraph{Baselines.}
We compare \SYS{} with four baselines.
\emph{vLLM-Marconi} uses vLLM~\cite{vllm} v0.27.1 in align mode with the turn-boundary checkpoint strategy of Marconi~\cite{pan2025marconi}, retaining prefix cache entries on the GPU.
vLLM does not yet support cache offloading for hybrid models.
We extend its offloading implementation to construct \emph{vLLM-CPU\textsuperscript{*}}, which offloads to CPU memory, and \emph{vLLM-Disk\textsuperscript{*}}, which offloads to both CPU memory and disk.
The asterisk denotes our extensions to vLLM. Both extensions follow the existing offloading framework, use vLLM's unified page organization for KV entries and recurrent states~\cite{zhang2025jenga}, and apply write-through caching and LRU eviction.
\emph{SGLang-HiCache} uses SGLang~\cite{sglang} v0.5.22 with HiCache~\cite{xie2026strata} enabled.
Its Unified Radix Cache~\cite{sglang2026unified} supports offloading both recurrent states and KV entries to CPU memory and disk.
We use its default cache partition, which allocates approximately half of the GPU cache capacity to each.
All systems use the same model, request streams, session concurrency limit, and GPU cache budget.
CPU and SSD budgets are matched across systems that use the corresponding tiers.

\paragraph{Workloads.}
We use two public datasets collected from real users of coding agents: Weka~\cite{fox2026weka} (\texttt{cc-traces-weka-042026}) and TraceLab~\cite{zhu2026tracelab}.
Weka contains multi-turn Claude Code sessions with input and output lengths, request timing, and identifiers for 64-token input blocks.
We use the Codex subset of TraceLab, which records per-round token usage, timing, and tool calls, and refer to it as TraceLab below.
We replay these traces~\cite{pan2026agentreplay} to evaluate cache reuse under real coding-agent workloads, including repeated prefix accesses and gaps between turns.
Besides, to isolate the effect of context mutation on cache reuse, we use microbenchmarks that vary mutation frequency while controlling context growth.

\paragraph{Evaluation methods.}
We set the maximum number of concurrently live sessions on the GPU to 24 to avoid memory thrashing under excessive concurrency~\cite{kang2026thunderagent,chen2026concur}.
Whenever the live-session count falls below this limit, we randomly select another session from the dataset and admit it for replay.
We choose the total number of sessions in each workload to target an overall replay duration of approximately three hours.
The Weka workload contains 3,784 requests from 125 sessions, and the TraceLab Codex workload contains 4,791 requests from 68 sessions.

\paragraph{Metrics.}
TTFT is measured from request arrival to the first generated token, including queueing, cache restoration, and recomputation.
Most requests append to existing context, for which retaining the preceding turn's terminal state is sufficient.
We therefore report p95 and p99 TTFT over all requests to expose the cost of less frequent reuse failures that aggregate latency can obscure.
We also report the token-level cache hit rate, defined as the total number of input tokens reused without recomputation divided by the total number of input tokens.
A KV prefix match alone does not count as reuse when the corresponding recurrent state must be recomputed.

\subsection{Real Agentic Trace Results}
\label{sec:eval-traces}

\begin{figure}[!t]
    \centering
    \includegraphics[width=\columnwidth]{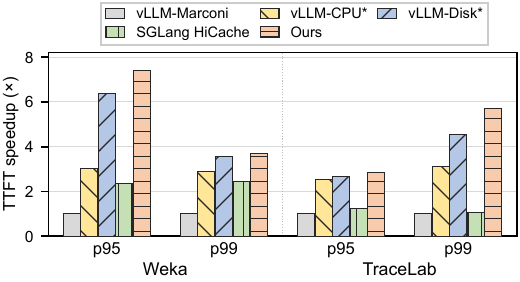}
    \caption{p95 and p99 TTFT speedup relative to vLLM-Marconi on Weka and TraceLab. Higher is better.}
    \label{fig:eval-trace-ttft}
\end{figure}

\paragraph{Tail latency.}
\SYS{} achieves the lowest p95 and p99 TTFT on both workloads, with speedups of up to 7.40$\times$ and 5.73$\times$, respectively, over vLLM-Marconi (Figure~\ref{fig:eval-trace-ttft}).
Each speedup divides the baseline's TTFT percentile by the corresponding percentile of the evaluated system.
Offloading alone substantially reduces tail latency by retaining prefixes beyond GPU capacity, but \SYS{} provides further gains over vLLM-Disk\textsuperscript{*}.
Additional capacity is therefore only part of the solution: the cache must also retain checkpoints that make returning prefixes reusable.
The incremental p99 benefit is smaller on Weka, where the slowest requests are less affected by the reuse improvements, than on TraceLab.
Meanwhile, p50 TTFT remains broadly comparable to vLLM-Disk\textsuperscript{*}, with a modest regression on Weka, indicating that the tail gains do not come with a large change in common-request latency.

HiCache has higher p95 and p99 TTFT than even vLLM-CPU\textsuperscript{*}, despite also using disk storage.
Its static partition reserves capacity for states that cannot be used to relieve KV pressure, allowing KV entries to be evicted before reuse.
The frequent KV evictions observed in these runs are consistent with this limitation and motivate sharing capacity across the two cache types.

\begin{figure}[!t]
    \centering
    \includegraphics[width=\columnwidth]{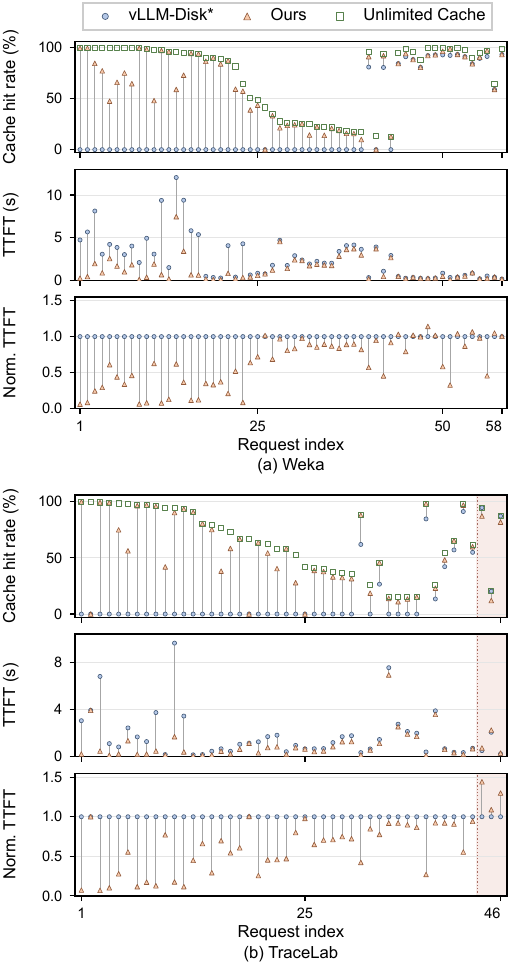}
    \caption{Cache hit rate, TTFT, and normalized TTFT for specific Weka and TraceLab requests. Unlimited Cache retains historical KV entries and recurrent states at every token position without eviction. Each workload uses the same request order across all panels. Shading marks requests with lower hit rates under \SYS{} than under vLLM-Disk\textsuperscript{*}.}
    \label{fig:eval-changed-requests}
\end{figure}

\paragraph{Per-request cache reuse.}
Figure~\ref{fig:eval-changed-requests} relates cache reuse to request latency.
Unlimited Cache is a reference that retains all previously computed KV entries and recurrent states at every token position without eviction.
Its hit rate is computed from reusable historical prefixes, subject to the cache's block granularity.
We focus on \emph{reuse-sensitive requests}: requests whose vLLM-Disk\textsuperscript{*} hit rate falls at least five percentage points below Unlimited Cache or whose hit rate differs by at least five percentage points between \SYS{} and vLLM-Disk\textsuperscript{*}.
Within each workload, all panels use the same order, sorted by decreasing hit-rate gap between Unlimited Cache and vLLM-Disk\textsuperscript{*}, with requests whose hit rate decreases under \SYS{} placed in the shaded region on the right.
Normalized TTFT divides each request's latency by its vLLM-Disk\textsuperscript{*} latency.

Higher cache reuse consistently accompanies lower TTFT among the requests that gain reuse.
Across these reuse-sensitive requests, mean TTFT decreases from 2.66 to 1.29 seconds on Weka and from 1.77 to 0.96 seconds on TraceLab.
Many gains come from recovering prefixes that vLLM-Disk\textsuperscript{*} misses entirely, avoiding repeated prefill computation.
The normalized panels show that this benefit also extends to requests whose absolute latencies are small.

Only a few requests lose reuse under \SYS{}, as shown by the shaded region.
These cases reflect the tradeoff made by our eviction policy: reclaiming a tail KV segment together with its states can release more memory at a limited expected recomputation cost.
If the reclaimed tail is reused, its reconstruction lowers the hit rate and adds latency.
The resulting absolute latency penalties are small relative to the multi-second misses avoided elsewhere, even when the normalized ratios rise visibly.
Thus, \SYS{} trades a small amount of reuse on a few requests for more space to retain useful prefixes across sessions.

Across these reuse-sensitive requests, \SYS{} recovers 87.0\% and 84.9\% of the reusable tokens under Unlimited Cache in Weka and TraceLab, respectively.
These ratios sum reused tokens across each cohort before dividing by the total under Unlimited Cache.
The remaining gap reflects the difference between indefinite retention and a finite cache budget: \SYS{} must sparsify checkpoints and reclaim inactive trajectories to make room for active sessions.
This tradeoff is visible in several requests where both systems miss despite substantial reuse under Unlimited Cache.
Those requests return after long idle intervals, as can occur during extended tool execution or waits for user input, by which time retaining their cache entries competes with ongoing work.

\begin{figure}[!t]
    \centering
    \includegraphics[width=\columnwidth]{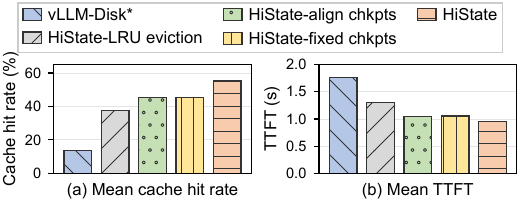}
    \caption{Ablation results on the same 46 TraceLab reuse-sensitive requests as Figure~\ref{fig:eval-changed-requests}. Variants replace one policy in \SYS{}. Bars show arithmetic means of per-request cache hit rates and TTFT.}
    \label{fig:eval-ablation}
\end{figure}

\paragraph{Ablation study.}
We examine the contributions of checkpoint admission and disk eviction on the TraceLab reuse-sensitive requests in Figure~\ref{fig:eval-changed-requests}.
We replace adaptive interval control with either vLLM's align checkpoint policy or a fixed interval of 32 blocks, retaining terminal checkpoints.
For eviction, we replace the disk eviction policy with vLLM-Disk\textsuperscript{*}'s file-level LRU policy.
Each variant retains the remaining mechanisms and the same cache budgets.

Figure~\ref{fig:eval-ablation} shows that both admission and eviction contribute to reuse and latency improvements.
LRU causes the largest loss in average hit rate and increases mean TTFT by 37\%, as useful prefixes are reclaimed before longer-gap sessions return.
Align checkpoints and fixed intervals leave fewer intermediate checkpoints near internal reuse positions, increasing recomputation and mean TTFT by about 9--10\%.
The two mechanisms are complementary: admission determines where computation can resume, while eviction preserves those opportunities across session gaps.

\subsection{Microbenchmarks}
\label{sec:eval-microbenchmarks}

\paragraph{Sensitivity to mutation frequency.}
To examine how context mutation frequency affects the benefits of retaining intermediate checkpoints, we compare \SYS{} with vLLM-Disk\textsuperscript{*} using synthetic microbenchmarks with mutation rates of 0\%, 2\%, 5\%, and 10\%.
We use the model and cache budgets in Section~\ref{sec:eval-setup}.
Each workload replays 24 concurrent sessions for 90 minutes.
Each session runs for 16 turns, with 8K new input tokens and 256 output tokens per turn, separated by three minutes.
Each mutation cuts the preceding prompt at a uniformly sampled internal 256-token boundary and appends the next 8K-token input without restoring the discarded suffix.

\begin{figure}[!t]
    \centering
    \includegraphics[width=\columnwidth]{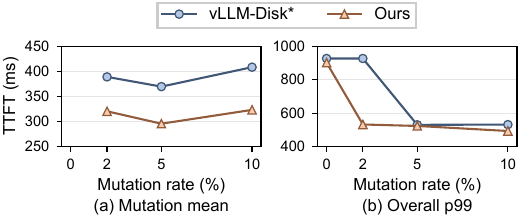}
    \caption{Mean TTFT of mutation requests after warmup (a) and overall p99 TTFT (b). Lower is better.}
    \label{fig:eval-mutation-frequency}
\end{figure}

Figure~\ref{fig:eval-mutation-frequency} reports mean TTFT for mutation requests in panel (a) and overall p99 TTFT in panel (b).
\SYS{} reduces mean TTFT for mutation requests by 18--21\%.
Its cache hierarchy retains checkpoints near internal reuse positions, allowing requests to resume without recomputing from earlier turn boundaries.
Cache hit rates for mutation requests improve by 7.6--7.7 percentage points across the nonzero settings.
As mutation frequency increases, more requests can benefit from these intermediate checkpoints.
The benefits also extend to the workload tail: overall p99 TTFT improves at every setting, with a 7.2\% reduction at the highest mutation rate.

\section{Related Work}
\label{sec:related-work}

\paragraph{LLM and agent serving.}
LLM serving systems improve resource utilization through iteration-level scheduling~\cite{yu2022orca}, paged KV allocation~\cite{vllm}, and prefill-decode disaggregation~\cite{zhong2024distserve}.
For agent applications, prior work exploits application dataflow and execution status to schedule requests~\cite{lin2024parrot,luo2025autellix,kang2026thunderagent,chen2026concur}, and manages cached contexts across tool interruptions and workflow steps~\cite{abhyankar2024infercept,pan2025kvflow,li2025continuum,pan2026scalesim,xia2026idleness,pan2026rrevict}.
SmoothAgent~\cite{pan2026smoothagent} further overlaps KV preparation for context changes with agent execution.
These works do not address recurrent checkpoint management for hybrid models. \SYS{} preserves the states needed to resume from matching KV prefixes, complementing their scheduling and prefetching optimizations.

\paragraph{Prefix cache management.}
Prior systems support prefix reuse through radix-tree indexing~\cite{sglang,pan2025fasttree}, hierarchical conversation caching~\cite{yu2025stateful,gao2024cost,xie2026strata}, and KV storage and transfer across inference engines~\cite{qin2025mooncake,liu2025lmcache}.
Hybrid models additionally require selecting which recurrent states to checkpoint.
Marconi~\cite{pan2025marconi} combines selective checkpoint admission with eviction based on recency and computation saved per unit of storage, while Sparse Prefix Caching~\cite{shirokikh2026sparse} optimizes checkpoint positions for a given budget and prefix-match distribution.
SGLang's Unified Radix Cache~\cite{sglang2026unified} supports heterogeneous cache components, hierarchical caching, and session-aware eviction.
\SYS{} jointly manages checkpoint placement and retention across tiers and concurrent agent sessions. Asymmetric tiering, adaptive interval control, and unified eviction improve cache reuse under context mutation and delayed access, reducing recomputation and TTFT.

\section{Conclusion}
\label{sec:conclusion}

We presented \SYS{}, a prefix cache management system for hybrid LLMs under agent context mutation and delayed reuse.
Its design exploits the fact that each prefix hit accesses all matching KV entries but only one recurrent checkpoint.
Asymmetric cache tiering prioritizes fast memory for KV prefixes and retains historical checkpoints on SSD, while adaptive interval control and unified eviction regulate checkpoint density and retention under resource pressure.
Together, these mechanisms balance checkpoint retention within trajectories with cache allocation across concurrent sessions.
Experiments on real agent traces and synthetic workloads demonstrate improved cache reuse and lower tail latency compared with baselines.

\bibliographystyle{mlsys2025}
\bibliography{references}

@article{lieber2024jamba,
  title={{Jamba: A hybrid Transformer-Mamba language model}},
  author={Opher Lieber and Barak Lenz and Hofit Bata and Gal Cohen and Jhonathan Osin and Itay Dalmedigos and Erez Safahi and Shaked Meirom and Yonatan Belinkov and Shai Shalev-Shwartz and Omri Abend and Raz Alon and Tomer Asida and Amir Bergman and Roman Glozman and Michael Gokhman and Avashalom Manevich and Nir Ratner and Noam Rozen and Erez Shwartz and Mor Zusman and Yoav Shoham},
  journal={arXiv preprint arXiv:2403.19887},
  year={2024}
}

@article{glorioso2024zamba,
  title={{Zamba: A compact 7B SSM hybrid model}},
  author={Glorioso, Paolo and Anthony, Quentin and Tokpanov, Yury and Whittington, James and Pilault, Jonathan and Ibrahim, Adam and Millidge, Beren},
  journal={arXiv preprint arXiv:2405.16712},
  year={2024}
}

@article{waleffe2024empirical,
  title={{An empirical study of Mamba-based language models}},
  author={Roger Waleffe and Wonmin Byeon and Duncan Riach and Brandon Norick and Vijay Korthikanti and Tri Dao and Albert Gu and Ali Hatamizadeh and Sudhakar Singh and Deepak Narayanan and Garvit Kulshreshtha and Vartika Singh and Jared Casper and Jan Kautz and Mohammad Shoeybi and Bryan Catanzaro},
  journal={arXiv preprint arXiv:2406.07887},
  year={2024}
}

@article{gu2023mamba,
  title={{Mamba: Linear-time sequence modeling with selective state spaces}},
  author={Gu, Albert and Dao, Tri},
  journal={arXiv preprint arXiv:2312.00752},
  year={2023}
}

@article{team2025kimi,
  title={{Kimi Linear: An expressive, efficient attention architecture}},
  author={{Kimi Team} and Yu Zhang and Zongyu Lin and Xingcheng Yao and Jiaxi Hu and Fanqing Meng and Chengyin Liu and Xin Men and Songlin Yang and Zhiyuan Li and Wentao Li and Enzhe Lu and Weizhou Liu and Yanru Chen and Weixin Xu and Longhui Yu and Yejie Wang and Yu Fan and Longguang Zhong and Enming Yuan and Dehao Zhang and Yizhi Zhang and T. Y. Liu and Haiming Wang and Shengjun Fang and Weiran He and Shaowei Liu and Yiwei Li and Jianlin Su and Jiezhong Qiu and Bo Pang and Junjie Yan and Zhejun Jiang and Weixiao Huang and Bohong Yin and Jiacheng You and Chu Wei and Zhengtao Wang and Chao Hong and Yutian Chen and Guanduo Chen and Yucheng Wang and Huabin Zheng and Feng Wang and Yibo Liu and Mengnan Dong and Zheng Zhang and Siyuan Pan and Wenhao Wu and Yuhao Wu and Longyu Guan and Jiawen Tao and Guohong Fu and Xinran Xu and Yuzhi Wang and Guokun Lai and Yuxin Wu and Xinyu Zhou and Zhilin Yang and Yulun Du},
  journal={arXiv preprint arXiv:2510.26692},
  year={2025}
}

@article{team2026kimi,
  title={{{Kimi K3}: Open frontier intelligence}},
  author={{Kimi Team} and Tongtong Bai and Yifan Bai and Yiping Bao and M. C. and Jianfeng Cai and Xinyuan Cai and Peizhou Cao and Yuxuan Cao and Ziwei Chai and Y. Charles and H. S. Che and Guanduo Chen and Guangyu Chen and Guanzheng Chen and Huarong Chen and Jia Chen and Jianlong Chen and Jun Chen and Kexin Chen and Peng Chen and Ruijue Chen and Wentao Chen and Xin Chen and Yang Chen and Yanru Chen and Yifei Chen and Yingjiang Chen and Yuankun Chen and Yujie Chen and Yutian Chen and Zhirong Chen and Dazhi Cheng and Yean Cheng and Jialei Cui and Jingbing Cui and Anqi Dai and Jiaqi Deng and Hao Ding and Rui Ding and Shaofeng Ding and Mengfan Dong and Mengnan Dong and Yuhao Dong and Yuxin Dong and Angang Du and Chenzhuang Du and Dikang Du and Jusen Du and Yulun Du and Yu Fan and Jing Feng and Qiulin Feng and Yichen Feng and Kelin Fu and Qiang Fu and Fuxuan Gao and Hongcheng Gao and Jingyue Gao and Tong Gao and Weijia Gao and Shangyi Geng and Jie Gong and Linhu Gong and Shengao Gong and Xiaochen Gong and Qizheng Gu and Yicheng Gu and Shuhao Guan and Haiqing Guo and Shiqi Guo and Xiang Guo and Zhengyan Guo and Beixi Hao and Wenxin Hao and Xiaoru Hao and Dailan He and Haotian He and Lehan He and Qi He and Weiran He and Xinran He and Xinyi He and Yibo He and Yunjia He and Chao Hong and Tiange Hong and Hao Hu and Jiaxi Hu and Ruikun Hu and Weiming Hu and Yangyang Hu and Zhenxing Hu and Liang Hua and Jinbin Huang and Ke Huang and Ruiyuan Huang and Siying Huang and Weixiao Huang and Yan Huang and Zhengjie Huang and Zhiqi Huang and Yulong Hui and Chaobo Jia and Yutong Jiang and Zhejun Jiang and Zuoyou Jiang and Wenyi Jin and Xinyi Jin and Yu Jing and Huanjun Kong and Guokun Lai and Aidi Li and Cheng Li and Chengyuan Li and Cong Li and Fang Li and Guanyu Li and Haoyang Li and Jia Li and Junxiong Li and Lei Li and Letian Li and Lincan Li and Weihong Li and Wentao Li and Xintong Li and Yang Li and Yishen Li and Yiwei Li and Yuxiao Li and Zhaowei Li and Zhaoxi Li and Zheming Li and Zhengxiao Li and Zhiyuan Li and Jiawei Lin and Xiaohan Lin and Yibo Lin and Zichao Lin and Ziyan Lin and Bill Liu and Boxiao Liu and Chuan Liu and Liang Liu and Shaowei Liu and Shudong Liu and Shuran Liu and Tianwei Liu and Weizhou Liu and Yangyang Liu and Yanming Liu and Yibo Liu and Yipeng Liu and Zhengying Liu and Zhiheng Liu and Enzhe Lu and Haoyu Lu and Linqiang Lu and Tingzhan Lu and Zhiyuan Lu and Aotian Luo and G. Luo and Junyu Luo and Yifan Luo and B. Lyu and Wenzhou Lyu and Shaoguang Mao and Yuan Mei and Xin Men and Minqing Ni and Yixuan Niu and Siyuan Pan and Shujun Peng and Zhangyang Qi and Ruoyu Qin and ZeChao Qin and Zeyu Qin and Haiquan Qiu and Jianxin Qiu and Jiezhong Qiu and Bowen Qu and Yuhao Qu and Zeyu Shang and Youbo Shao and Han Shen and Jincheng Shi and Juanfeng Shi and Lidong Shi and Shengyuan Shi and Wingchun Siu and Pengwei Song and Xiaoxi Song and Jianlin Su and Yunfeng Su and Zhaochen Su and Lin Sui and Jingsong Sun and Junyao Sun and Shaoning Sun and Shuzhe Sun and Tongyu Sun and Yujun Sun and Yunpeng Tai and Chuning Tang and Heyi Tang and Sirui Tang and Zecheng Tang and Chaoran Tian and Rongpeng Tian and Yu Tian and Wei Tu and Chensi Wang and Chuang Wang and Chunjie Wang and Dinglu Wang and Feng Wang and Hailong Wang and Haiming Wang and Hao Wang and Hao Wang and Huaqing Wang and Hui Wang and Jiayi Wang and Jinglong Wang and Jinhong Wang and Jiuzheng Wang and Linian Wang and Shaobo Wang and Shenzhi Wang and Shuyi Wang and Si Wang and Siyuan Wang and Tianfu Wang and Wenjue Wang and Xingran Wang and Xinmei Wang and Xinyuan Wang and Xusheng Wang and Yalin Wang and Yangkun Wang and Yao Wang and Yaoyu Wang and Yejie Wang and Yiqin Wang and Yucheng Wang and Yuzhi Wang and Zhaoji Wang and Zhaowei Wang and Zhengtao Wang and Zhenhao Wang and Zhongsheng Wang and Zifan Wang and Chu Wei and Ming Wei and Shouxin Wei and Zichen Wen and Fan Wu and Haoning Wu and Rucong Wu and Wenhao Wu and Xiaoxue Wu and Yingcong Wu and Yongqi Wu and Yuxin Wu and Zijian Wu and Xinglang Xian and Chenxuan Xiang and Yuye Xiang and Bocheng Xiao and Chenjun Xiao and Xin Xiao and Jin Xie and Xiaotong Xie and Yifeng Xie and Zhe Xie and Bowei Xing and Yiming Xiong and Baosheng Xu and Boyu Xu and Jiale Xu and Jianfan Xu and Jing Xu and Jinjing Xu and L. H. Xu and Qingtao Xu and Shuyao Xu and Suting Xu and Tiantian Xu and Tianxiang Xu and Weixin Xu and Xinran Xu and Yangchuan Xu and Ye Xu and Yueni Xu and Ziyao Xu and Haonan Xue and Junjie Yan and Yaoyao Yan and Fan Yang and Guangyao Yang and Hao Yang and Junwei Yang and Ruoyu Yang and Wenjie Yang and Xiaofei Yang and Xinyu Yang and Yi Yang and Yiling Yang and Ying Yang and Yuchen Yang and Zhen Yang and Zhilin Yang and Zian Yang and Zuhao Yang and Haotian Yao and Dan Ye and Haoran Ye and Wenjie Ye and Zhanbo Ye and Bohong Yin and Haoxiang Yin and Xietong Yin and Chengzhen Yu and Haozhen Yu and Longhui Yu and Shengnan Yu and Shuying Yu and Tianxiang Yu and Enming Yuan and Mengjie Yuan and Tongtian Yue and Wei Yue and Yang Yue and Dunyuan Zha and Haobing Zhan and B. H. Zhang and Dehao Zhang and Fei Zhang and Hao Zhang and Haoyuan Zhang and Huanyu Zhang and Jiapei Zhang and Jiaxuan Zhang and Jin Zhang and Kaiyi Zhang and Miaozhen Zhang and Puqi Zhang and Qinglei Zhang and Rong Zhang and Rui Zhang and Shaoshuai Zhang and Shiyi Zhang and Xiaobin Zhang and Xiaoyun Zhang and Y. Zhang and Yangkun Zhang and Ye Zhang and Yichi Zhang and Yikun Zhang and Yizhi Zhang and Yongting Zhang and Yu Zhang and Yutao Zhang and Yutong Zhang and Zheng Zhang and Zijing Zhang and Bin Zhao and Chenguang Zhao and Feifan Zhao and Jinglun Zhao and Jinxiang Zhao and Shuai Zhao and Wenshuo Zhao and Xiangyu Zhao and Xuanle Zhao and Yikai Zhao and Zijia Zhao and Haozhi Zheng and Huabin Zheng and Ruihan Zheng and Shaojie Zheng and Tengyang Zheng and Haofeng Zhong and Lei Zhong and Longguang Zhong and M. Zhou and Qiankang Zhou and Runjie Zhou and Ruozhang Zhou and Xinyu Zhou and Yiqiao Zhou and Zaida Zhou and Jinguo Zhu and Liya Zhu and Xinhao Zhu and Yangjunfeng Zhu and Yuxuan Zhu and Zhen Zhu and Chen Zhuang and Weiyu Zhuang and Xinxing Zu},
  journal={arXiv preprint arXiv:2607.24653},
  year={2026}
}

@misc{qwen2026qwen35,
  title ={{{Qwen3.5}: Towards Native Multimodal Agents}},
  author = {{Qwen Team}},
  month = feb,
  year = {2026},
  howpublished = {\url{https://qwen.ai/blog?id=qwen3.5}}
}

@inproceedings{vllm,
  title={{Efficient memory management for large language model serving with PagedAttention}},
  author={Kwon, Woosuk and Li, Zhuohan and Zhuang, Siyuan and Sheng, Ying and Zheng, Lianmin and Yu, Cody Hao and Gonzalez, Joseph and Zhang, Hao and Stoica, Ion},
  booktitle={Proceedings of the 29th symposium on operating systems principles},
  pages={611--626},
  year={2023}
}

@article{pan2025marconi,
  title={{Marconi: Prefix caching for the era of hybrid LLMs}},
  author={Pan, Rui and Wang, Zhuang and Jia, Zhen and Karakus, Can and Zancato, Luca and Dao, Tri and Wang, Yida and Netravali, Ravi},
  journal={Proceedings of Machine Learning and Systems},
  volume={7},
  year={2025}
}

@inproceedings{katharopoulos2020transformers,
  title={{Transformers are RNNs: Fast autoregressive transformers with linear attention}},
  author={Katharopoulos, Angelos and Vyas, Apoorv and Pappas, Nikolaos and Fleuret, Fran{\c{c}}ois},
  booktitle={International conference on machine learning},
  pages={5156--5165},
  year={2020},
  organization={PMLR}
}

@article{sglang,
  title={{SGLang: Efficient execution of structured language model programs}},
  author       = {Lianmin Zheng and
                  Liangsheng Yin and
                  Zhiqiang Xie and
                  Chuyue Sun and
                  Jeff Huang and
                  Cody Hao Yu and
                  Shiyi Cao and
                  Christos Kozyrakis and
                  Ion Stoica and
                  Joseph E. Gonzalez and
                  Clark Barrett and
                  Ying Sheng},
  journal={Advances in Neural Information Processing Systems},
  volume={37},
  pages={62557--62583},
  year={2024}
}

@article{pan2026smoothagent,
  title={{{SmoothAgent}: Efficient Long-Horizon LLM-Based Agent Serving with Lookahead Context Engineering}},
  author={Pan, Zaifeng and Wang, Qianxu and Hu, Zhengding and Chen, Chang and Guan, Yue and Zhou, Yanbo and Swanson, Steven and Ding, Yufei},
  journal={arXiv preprint arXiv:2607.00151},
  year={2026}
}

@article{shirokikh2026sparse,
  title={{Sparse Prefix Caching for Hybrid and Recurrent LLM Serving}},
  author={Shirokikh, Mikhail and Nikolenko, Sergey},
  journal={arXiv preprint arXiv:2605.05219},
  year={2026}
}

@misc{sglang2026unified,
  title ={{Unified Radix Cache in {SGLang}: One Tree for Hybrid Model Prefix Caching}},
  author = {Zhang, Yichi and {SGLang Team}},
  year = {2026},
  howpublished = {\url{https://www.sglang.io/blog/unified-radix-cache}},
}

@article{zhu2026tracelab,
  title={{TraceLab: Characterizing Coding Agent Workloads for LLM Serving}},
  author={Zhu, Kan and Jacob, Mathew and Ma, Chenxi and Pan, Yi and Wang, Stephanie and Krishnamurthy, Arvind and Kasikci, Baris},
  journal={arXiv preprint arXiv:2606.30560},
  year={2026}
}

@misc{fox2026weka,
  author = {{SemiAnalysis}},
  title ={{{CC Traces}: {Weka} (April 2026)}},
  year = {2026},
  howpublished = {\url{https://huggingface.co/datasets/semianalysisai/cc-traces-weka-042026}},
  note = {Published by SemiAnalysis}
}

@misc{ji2025manus,
  author = {Ji, Yichao},
  title ={{Context Engineering for {AI} Agents: Lessons from Building Manus}},
  howpublished = {\url{https://manus.im/blog/Context-Engineering-for-AI-Agents-Lessons-from-Building-Manus}},
  year = {2025}
}

@misc{langchain_context_engineering,
  author = {LangChain},
  title ={{Context Engineering}},
  howpublished = {\url{https://blog.langchain.com/context-engineering-for-agents}},
  year = {2025}
}

@misc{context_for_deepagents,
  author = {Chester Curme and Mason Daugherty},
  title ={{Context Management for Deep Agents}},
  howpublished = {\url{https://blog.langchain.com/context-management-for-deepagents}},
  year = {2026}
}

@misc{anthropic_context_engineering,
  author = {Anthropic},
  title ={{Effective context engineering for AI agents}},
  howpublished = {\url{https://www.anthropic.com/engineering/effective-context-engineering-for-ai-agents}},
  year = {2025}
}

@inproceedings{yao2023react,
  title={{ReAct: Synergizing reasoning and acting in language models}},
  author={Yao, Shunyu and Zhao, Jeffrey and Yu, Dian and Du, Nan and Shafran, Izhak and Narasimhan, Karthik and Cao, Yuan},
  booktitle={International Conference on Learning Representations (ICLR)},
  year={2023}
}

@article{yang2024swe,
  title={{SWE-agent: Agent-computer interfaces enable automated software engineering}},
  author={Yang, John and Jimenez, Carlos E and Wettig, Alexander and Lieret, Kilian and Yao, Shunyu and Narasimhan, Karthik and Press, Ofir},
  journal={Advances in Neural Information Processing Systems},
  volume={37},
  pages={50528--50652},
  year={2024}
}

@misc{openclaw,
  author = {Steinberger, Peter and {OpenClaw Community}},
  title ={{{OpenClaw}: Personal AI Assistant}},
  howpublished = {\url{https://openclaw.ai}},
  year = {2026}
}

@misc{langchain,
  author = {LangChain},
  title ={{{LangChain}: The agent engineering platform.}},
  howpublished = {\url{https://www.langchain.com}},
  year = {2026}
}

@article{pan2025kvflow,
  title={{{KVFlow}: Efficient Prefix Caching for Accelerating LLM-Based Multi-Agent Workflows}},
  author={Pan, Zaifeng and Patel, Ajjkumar and Hu, Zhengding and Shen, Yipeng and Guan, Yue and Li, Wan-Lu and Qin, Lianhui and Wang, Yida and Ding, Yufei},
  journal={arXiv preprint arXiv:2507.07400},
  year={2025}
}

@inproceedings{gao2024cost,
  title={{Cost-Efficient large language model serving for multi-turn conversations with {CachedAttention}}},
  author={Gao, Bin and He, Zhuomin and Sharma, Puru and Kang, Qingxuan and Jevdjic, Djordje and Deng, Junbo and Yang, Xingkun and Yu, Zhou and Zuo, Pengfei},
  booktitle={2024 USENIX Annual Technical Conference (USENIX ATC 24)},
  pages={111--126},
  year={2024}
}

@article{liu2025lmcache,
  title={{LMCache: An efficient KV cache layer for enterprise-scale LLM inference}},
  author={Liu, Yuhan and Cheng, Yihua and Yao, Jiayi and An, Yuwei and Chen, Xiaokun and Feng, Shaoting and Huang, Yuyang and Shen, Samuel and Zhang, Rui and Du, Kuntai and Jiang, Junchen},
  journal={arXiv preprint arXiv:2510.09665},
  year={2025}
}

@inproceedings{yu2025stateful,
  title={{Stateful large language model serving with Pensieve}},
  author={Yu, Lingfan and Lin, Jinkun and Li, Jinyang},
  booktitle={Proceedings of the Twentieth European Conference on Computer Systems},
  pages={144--158},
  year={2025}
}

@inproceedings{qin2025mooncake,
  title={{Mooncake: Trading more storage for less computation—a {KVCache-centric} architecture for serving {LLM} chatbot}},
  author={Qin, Ruoyu and Li, Zheming and He, Weiran and Cui, Jialei and Ren, Feng and Zhang, Mingxing and Wu, Yongwei and Zheng, Weimin and Xu, Xinran},
  booktitle={23rd USENIX Conference on File and Storage Technologies (FAST 25)},
  pages={155--170},
  year={2025}
}

@misc{codex,
  author = {{OpenAI}},
  title ={{{Codex}: Open-source coding agent}},
  year = {2026},
  howpublished = {\url{https://github.com/openai/codex}}
}

@misc{vllm_prefix_caching,
  author = {{vLLM Team}},
  title ={{{vLLM} Automatic Prefix Caching}},
  year = {2026},
  howpublished = {\url{https://docs.vllm.ai/en/latest/design/prefix_caching/}}
}

@inproceedings{cao1997greedydual,
  author = {Cao, Pei and Irani, Sandy},
  title ={{Cost-Aware {WWW} Proxy Caching Algorithms}},
  booktitle = {USENIX Symposium on Internet Technologies and Systems (USITS 97)},
  year = {1997},
  pages = {193--206},
  publisher = {USENIX Association},
  url = {https://www.usenix.org/conference/usits-97/cost-aware-www-proxy-caching-algorithms}
}

@inproceedings{yu2022orca,
  author = {Yu, Gyeong-In and Jeong, Joo Seong and Kim, Geon-Woo and Kim, Soojeong and Chun, Byung-Gon},
  title ={{{Orca}: A Distributed Serving System for {Transformer-Based} Generative Models}},
  booktitle = {16th USENIX Symposium on Operating Systems Design and Implementation (OSDI 22)},
  pages = {521--538},
  year = {2022},
  publisher = {USENIX Association},
  url = {https://www.usenix.org/conference/osdi22/presentation/yu}
}

@inproceedings{zhong2024distserve,
  author = {Zhong, Yinmin and Liu, Shengyu and Chen, Junda and Hu, Jianbo and Zhu, Yibo and Liu, Xuanzhe and Jin, Xin and Zhang, Hao},
  title ={{{DistServe}: Disaggregating Prefill and Decoding for Goodput-optimized Large Language Model Serving}},
  booktitle = {18th USENIX Symposium on Operating Systems Design and Implementation (OSDI 24)},
  pages = {193--210},
  year = {2024},
  publisher = {USENIX Association},
  url = {https://www.usenix.org/conference/osdi24/presentation/zhong-yinmin}
}

@inproceedings{lin2024parrot,
  author = {Lin, Chaofan and Han, Zhenhua and Zhang, Chengruidong and Yang, Yuqing and Yang, Fan and Chen, Chen and Qiu, Lili},
  title ={{{Parrot}: Efficient Serving of {LLM-based} Applications with Semantic Variable}},
  booktitle = {18th USENIX Symposium on Operating Systems Design and Implementation (OSDI 24)},
  pages = {929--945},
  year = {2024},
  publisher = {USENIX Association},
  url = {https://www.usenix.org/conference/osdi24/presentation/lin-chaofan}
}

@article{luo2025autellix,
  title ={{{Autellix}: An Efficient Serving Engine for {LLM} Agents as General Programs}},
  author = {Luo, Michael and Shi, Xiaoxiang and Cai, Colin and Zhang, Tianjun and Wong, Justin and Wang, Yichuan and Wang, Chi and Huang, Yanping and Chen, Zhifeng and Gonzalez, Joseph E. and Stoica, Ion},
  journal = {arXiv preprint arXiv:2502.13965},
  year = {2025}
}

@article{abhyankar2024infercept,
  title ={{{InferCept}: Efficient Intercept Support for Augmented Large Language Model Inference}},
  author = {Abhyankar, Reyna and He, Zijian and Srivatsa, Vikranth and Zhang, Hao and Zhang, Yiying},
  journal = {arXiv preprint arXiv:2402.01869},
  year = {2024}
}

@inproceedings{xie2026strata,
  title={{Strata: Hierarchical context caching for long context language model serving}},
  author={Xie, Zhiqiang and Xu, Ziyi and Zhao, Mark and An, Yuwei and Mailthody, Vikram Sharma and Mahlke, Scott and Garland, Michael and Kozyrakis, Christos},
  booktitle={20th USENIX Symposium on Operating Systems Design and Implementation (OSDI 26)},
  pages={1--16},
  year={2026}
}

@inproceedings{zhang2025jenga,
  title={{Jenga: Effective memory management for serving LLM with heterogeneity}},
  author = {Zhang, Chen and Du, Kuntai and Liu, Shu and Kwon, Woosuk and Mo, Xiangxi and Wang, Yufeng and Liu, Xiaoxuan and You, Kaichao and Li, Zhuohan and Long, Mingsheng and Zhai, Jidong and Gonzalez, Joseph and Stoica, Ion},
  booktitle={Proceedings of the ACM SIGOPS 31st Symposium on Operating Systems Principles},
  pages={446--461},
  year={2025}
}

@article{kang2026thunderagent,
  title={{ThunderAgent: A simple, fast and program-aware agentic inference system}},
  author={Kang, Hao and Li, Ziyang and Xu, Weili and Yang, Xinyu and Chen, Yinfang and Wang, Junxiong and Chen, Beidi and Krishna, Tushar and Xu, Chenfeng and Arora, Simran},
  journal={arXiv preprint arXiv:2602.13692},
  year={2026}
}

@article{chen2026concur,
  title={{CONCUR: High-throughput agentic batch inference of LLM via congestion-based concurrency control}},
  author={Chen, Qiaoling and Ye, Zhisheng and Tang, Tian and Sun, Peng and Tian, Boyu and Wang, Guoteng and Li, Shenggui and Wen, Yonggang and Han, Zhenhua and Zhang, Tianwei},
  journal={arXiv preprint arXiv:2601.22705},
  year={2026}
}

@article{li2025continuum,
  title={{Continuum: Efficient and robust multi-turn LLM agent scheduling with KV cache time-to-live}},
  author={Hanchen Li and Runyuan He and Qiuyang Mang and Qizheng Zhang and Huanzhi Mao and Xiaokun Chen and Hangrui Zhou and Huanchen Zhang and Alvin Cheung and Joseph Gonzalez and Ion Stoica},
  journal={arXiv preprint arXiv:2511.02230},
  year={2025}
}

@article{xia2026idleness,
  title={{Idleness is Relative: Exploiting Tool-Call Idle Windows for Offloading in Agentic Systems with MORI}},
  author={Xia, Tian and Li, Hanchen and Li, Zhifei and Chen, Xiaokun and Kang, Hao and Qiao, Yifan and Xu, Yi and Stoica, Ion},
  journal={arXiv preprint arXiv:2606.00866},
  year={2026}
}

@article{pan2026scalesim,
  title={{ScaleSim: Serving Large-Scale Multi-Agent Simulation with Invocation Distance-Based Memory Management}},
  author={Pan, Zaifeng and Shen, Yipeng and Hu, Zhengding and Wang, Zhuang and Manocha, Aninda and Wang, Zheng and Yu, Zhongkai and Guan, Yue and Ding, Yufei},
  journal={arXiv preprint arXiv:2601.21473},
  year={2026}
}

@inproceedings{pan2025fasttree,
  title={{FastTree: Optimizing Attention Kernel and Runtime for Tree-Structured LLM Inference}},
  author={Pan, Zaifeng and Ding, Yitong and Guan, Yue and Wang, Zheng and Yu, Zhongkai and Tang, Xulong and Wang, Yida and Ding, Yufei},
  booktitle={Proceedings of Machine Learning and Systems},
  volume={7},
  year={2025}
}

@article{pan2026rrevict,
  title={{RR-Evict: Fine-Grained Prefix Cache Eviction beyond LRU for Agentic LLM Serving}},
  author={Pan, Zaifeng and Wu, Chris and Hu, Zhengding and Qiang, Xinwei and Yu, Zhongkai and Ding, Yufei},
  journal={arXiv preprint arXiv:2609.32278},
  year={2026}
}

@article{pan2026agentreplay,
  title={{AgentReplay: Token-Wise Trace Replay Is Essential for Fair Serving System Performance Benchmarking}},
  author={Pan, Zaifeng and Wang, Michael and Wu, Chris and Hu, Zhengding and Qiang, Xinwei and Yu, Zhongkai and Ding, Yufei},
  journal={arXiv preprint arXiv:2609.32283},
  year={2026}
}

\section*{AI Usage Disclosure}
We used OpenAI Codex and ChatGPT to discuss research ideas, assist with manuscript writing, and prepare figures.
We used Claude Code to assist with code development and experiments.
The authors take full responsibility for the correctness, originality, and integrity of the paper.

\end{document}